# Top-Quark Mass Data and the Sum of Quasi-Degenerate Neutrino Masses

(One small electroweak-bound ε-parameter organizes elementary particle 3-flavor phenomenology)

E. M. Lipmanov

40 Wallingford Road # 272, Brighton MA 02135, USA

**Abstract**

The absolute neutrino masses and type of neutrino mass hierarchy are among the main problems in neutrino physics. Top-quark mass is another topical problem in particle physics. These problems extend the old puzzle of electron-muon mass ratio close to the fine structure α-constant, which is still not solved by known theory. Here I continue the search for a general flavor pattern that may incorporate these problems. Relations between neutrino/electron and electron/top-quark pole mass ratios $3m_\nu/m_e \cong \pi\alpha_o^3 \cong m_e/3m_t$, with $\alpha_o$ a small empirical parameter close to the fine structure constant α, are obtained from supposition that realistic elementary particle dimensionless bare flavor quantities are small deviated (measured by universal parameter $\varepsilon = \sqrt{\alpha_o}$) from the values of a stated flavor pattern (at $\varepsilon = 0$) and experimental data hints. With the world average t-quark mass data the sum of QD-neutrino masses is estimated $\Sigma m_\nu = (0.50 \pm 0.003)$ eV in agreement with cosmological constraints and known QD-neutrino mass estimations from data on neutrino oscillation mass-squared differences.

# 1. **Introduction. Quasi-Degenerate neutrinos**

If the lepton and quark masses are parts of one whole particle mass system, one may expect uniting regularities between lepton and quark mass ratios, mixing angles and CP-violating phases on the level of low energy data.

The discussed in the literature empirical quark-neutrino complementarity mixing angle relations [1] present an example of such unity[1]. As shown in publication [2], those quark-neutrino complementarity relations are a part of physical flavor pattern which is guided by idea that at main tree approximation the relations between actual elementary particle flavor quantities quantitatively follow from quadratic hierarchy flavor rule, Dirac-Majorana deviation from mass-degeneracy (DMD) duality rule and *one small universal empirical parameter* $\sqrt{\alpha_o}$. Neutrino DMD-quantities are described by positive powers of that small parameter $\sqrt{\alpha_o}$ while the DMD-quantities of Dirac particles (CL and quarks) are described by negative powers of $\sqrt{\alpha_o}$. If experimentally confirmed that neutrinos are Majorana particles and of quasi-degenerate type, that condition means Dirac-Majorana DMD-duality for known elementary particles.

The mentioned above small empirical parameter is approximately given by

$$\alpha_o \cong e^{-5}, \quad \varepsilon \cong \sqrt{\alpha_o} = e^{-5/2}. \tag{1}$$

After emergence of the finite value (1) of $\alpha_o$-parameter (i.e. after the small shift of the parameter from value

---

[1] See also [6].

$\alpha_o \cong 0$ to the finite value (1)), relations for CL and QD-neutrino mass ratios are obtained [3] from known data[2]

$$(m_\mu/m_\tau) \cong \varepsilon/\sqrt{2}, \quad (m_e/m_\mu) \cong \varepsilon^2/\sqrt{2}, \quad (m_e/m_\tau) \cong \varepsilon^3/2, \qquad (2)$$

$$(m_3/m_2) \cong \exp(r), \quad (m_2/m_1) \cong \exp(r^2), \qquad (3)$$

where

$$r = \Delta m^2_{sol}/\Delta m^2_{atm} \cong 5\alpha_o \cong 1/30 \qquad (4)$$

is the neutrino oscillation solar-atmospheric hierarchy parameter, and normal ordering is chosen for neutrino masses.

Besides generating mass ratios, the small change of the parameter from $\varepsilon = 0$ to (1) produces realistic values of neutrino and quark mixing angles [2]:

$$\mathrm{Cos}^2 2\theta_{12} \cong \mathrm{Sin}^2 2\theta_c \cong (2\sqrt{2})(m_\mu/m_\tau) \cong 2\varepsilon, \qquad (5)$$

$$\mathrm{Cos}^2 2\theta_{23} \cong \mathrm{Sin}^2 2\theta' \cong (2\sqrt{2})(m_e/m_\mu) \cong 2\varepsilon^2. \qquad (6)$$

Here $\theta_{12}$ and $\theta_{23}$ are the solar and atmospheric neutrino oscillation large mixing angles, $\theta_c$ and $\theta'$ are the quark Cabibbo angle and next to it mixing one. It should be noted that the connections (5) and (6) incorporate approximately (comp. [12]) the known neutrino tribimaximal mixing pattern and the suggested [1] quark-neutrino complementarity mixing angle relations.

The two sequences (5) and (6) are connected by the considered [2] quadratic DMD-hierarchy rule for dimensionless flavor quantities, e.g. in case of neutrino masses

$$[(m_3/m_2)^2 - 1]^2 \cong 2[(m_2/m_1)^2 - 1].$$

---

[2] The universal parameter $\varepsilon$ is of empirical origin. From muon-electron PDG mass-ratio data [4] and relations (2) it follows $\varepsilon \cong 0.083$, $\log \varepsilon \cong -2.49 \cong -5/2$ to within ~0.3%.

That flavor hierarchy, duality and universal ε-parameter rules are supported by phenomenological solution of the problem of large neutrino mixing angles versus small quark ones, experimental data for CL mass ratios (2), prediction of QD-neutrino masses, and small solar-atmospheric neutrino oscillation hierarchy parameter $r \ll 1$ (4).

From relations (3) follows the absolute neutrino mass scale

$$(m_1)_{b.f.} \cong (\Delta m^2_{sol}/2r^2)^{1/2} \cong 0.18 \text{ eV}. \quad (7)$$

The best fit value [5, 11, 7] for solar mass-squared difference $(\Delta m^2_{sol})_{b.f.} \cong 7.6 \times 10^{-5}$ eV$^2$, and the choice $r \cong 1/30$, are used in (7). With 3σ ranges [5] for oscillation data $\Delta m^2_{sol} \cong (7.1 - 8.3) \times 10^{-5}$ eV$^2$, $r \cong (0.027 - 0.040)$, the estimation for absolute QD-neutrino mass scale is given by:

$$m_2 \cong (0.15 - 0.24) \text{ eV}. \quad (8)$$

I apply here the same *semi-empirical* approach to search for new manifestations of quark-lepton low energy mass matrix unity. As shown below, that approach points to a seesaw-like relation between small QD-neutrino and large top-quark masses, which confines the whole particle mass spectrum to known three quark-lepton generations.

## 2. ε = 0 approximation for particle 3-flavor pattern

New quantitative physical idea is that the realistic dimensionless particle flavor quantities (mass ratios, mixing angles, CP-violating phases) are small deviated from their values at 'background pattern' defined by three finite parameters – one dimensional mass parameter equal to the electron mass $m_e$, special angle 45° equal to maximal neutrino mixing angle and equal 1 mass ratios of degenerate maximally mixed massless[3] neutrinos, with all other particle mass ratios, quark mixing angles and CP-violating phases equal zero.

So in essence, elementary particle flavor pattern is described in terms of dimensionless physical quantities – angles, phases and different particle-electron mass ratios.

The zero $\varepsilon$-approximation of elementary particle flavor pattern with quark (q) and neutrino ($\nu$) mixing matrices is given by

$$[m_e \neq 0, \ m_\nu \cong 0, \ (m_\mu, \ m_\tau, \ m_q) \cong \infty],$$

$$m_\nu/m_e \cong 0, \ m_e/(m_\mu, \ m_\tau, \ m_q) \cong 0, \quad (9)$$

$$\begin{pmatrix} 1 & 0 & 0 \\ 0 & 1 & 0 \\ 0 & 0 & 1 \end{pmatrix}_q, \quad \begin{pmatrix} 1/\sqrt{2} & 1/\sqrt{2} & 0 \\ -1/2 & 1/2 & 1/\sqrt{2} \\ 1/2 & -1/2 & 1/\sqrt{2} \end{pmatrix}_\nu. \quad (10)$$

Small deviations from that background particle flavor pattern are parameterized by the small dimensionless universal parameter $\varepsilon$ (1) (see details in the lists (9) and (11) for neutrino and quark mixing angles respectively in ref.[2b])[4].

The background flavor pattern is the zero approximation for the expansions in $\varepsilon$-powers of the realistic flavor pattern of particle masses and mixings, with $m_e$ being the one primary mass parameter needed for dimensional definition of all other elementary particle masses. That background pattern (9),(10) describes the main features of 'small' versus 'large' quantities of experimental flavor data - in particular the connections between small quark mixing, large neutrino mixing, degenerate neutrinos and extraordinary small neutrino masses (in comparison with known Dirac particle masses) - as *primary* conditions of particle flavor physics with realistic flavor pattern as small deviated from this primary one.

---

[3] Strict conditions $\varepsilon = 0$, massless degenerate neutrinos and equal zero particle mass ratios are unattainable [2]. Note that the used here conditions of the type $m_q \cong \infty$, or $\varepsilon \cong 0$, are 'physical' conditions and mean 'finite, but unrestricted large, or close to zero'.

[4] On connection between small quark mixing, large neutrino mixing and QD-neutrino type in a gauge model see [18]. I would like to thank R. N. Mohapatra and M. K. Parida for the interest in my work [2].

If the neutrinos are Majorana particles, the special role of electron mass is that $m_e$ is a border-point between the upper Dirac particles (charged leptons and quarks) and lower Majorana particles (neutrinos).

## 3. **<u>Realistic one-parameter particle flavor pattern by small deviation from the $\varepsilon = 0$ approximation</u>**

The realistic elementary particle mass pattern should be approximately[5] described by three primary finite parameters: two dimensionless constants $\varepsilon$ and $45°$ and one dimensional mass constant $m_e$. Dual Dirac-Majorana particle relations are described by different discrete integer powers of the small and large factors '$\varepsilon$' and '$1/\varepsilon$': $\varepsilon$-powers for Majorana

---

[5] Since there is no finite field theory yet (but the data values are finite), one cannot quantitatively estimate the contributions of radiative SM corrections to given bare masses and mass ratios. But the approximate agreement with experimental data of the whole *system* of $\varepsilon$-representations of particle mass ratio and mixing angle bare values shows that it is natural to consider that corrections relatively small (calculable in a finite theory).

particles have positive signs while the ones for Dirac particles - negative signs.

Consider neutrino/electron and electron/top-quark mass *ratios* in the form[6]:

$$m_\nu/m_e = a\,\alpha_o^x, \quad m_t/m_e = (b\,\alpha_o^y), \quad x > 0, \quad y < 0, \qquad (11)$$

where (a, x) and (b, y) are four unknowns and $m_\nu$ is the average QD-neutrino mass $m_\nu = (m_1 + m_2 + m_3)/3 \equiv \Sigma m_\nu/3$. Since $\alpha_o$ is a very small parameter, an effective search for the exponents x and y in (11) can be made by comparison with empirical data[7]. The semi-empirical estimations of absolute QD-neutrino mass suggest x = 3, a = (0.96 – 1.53) from 3σ-ranges (8), and a = 1.15 from best fit estimation (7). Similar inference can be made from t-quark PDG data [4]

$$m_t^{exp} = (174.2 \pm 3.3)\text{ GeV}. \qquad (12)$$

These data suggest y = − 3, b ≅ 1/9.59 from central value in (12). If instead, one chooses for the coefficient b the close value

$$b = 1/3\pi, \qquad (13)$$

the second relation in (11) reads

$$m_e/3m_t \cong \pi\,\alpha_o^3. \qquad (14)$$

The left side of relation (14) has physical meaning – it is the ratio of electron mass to the *sum* of three color degenerate top-quark masses. The semi-empirical relation (14) indicates equal numbers (symmetry) of quark colors, coefficient '3' at the left side, and quark flavors, $\alpha_o$-power '3' on the right.

---

[6] It is supposed that coefficients 'a' and 'b' are not very large.

[7] Order-one parts of the coefficients in different considered below relations are expressed through small powers of only two numbers '2' and 'π', comp. [13].

The emerged sum of degenerate top-quark masses in (14) is a relevant indication from the experimental data - color symmetry of all physical quantities and relations is the main consistency condition for quarks.

From (14) the top-mass,

$$m_t \cong m_e/(3\pi\alpha_o^3) \cong 177.2 \text{ GeV}, \qquad (15)$$

is in good agreement with experimental data (12), to within ~1 S.D.

Consider now the recent new world average preliminary evaluation of the top-quark mass [8]

$$m_t^{exp} = (172.6 \pm 1.4) \text{ GeV}. \qquad (16)$$

It differs from the PDG data (12) by ~1 S.D., but is more accurate (~0.8%). With the central experimental value (16), one finds $[(m_e/3m_t^{ctr})/\pi\alpha_o^3 - 1]/\alpha_o \cong 3.992$. So, the new value (16) for the three-fold color degenerate t-quark mass may be correctly expressed in terms of the universal parameter $\alpha_o$ by a simple formula including a second term in $\alpha_o$-expansion (comp.(14):

$$m_e/3m_t = \pi\alpha_o^3(1 + 4\alpha_o), \quad m_t \cong 172.59 \text{ GeV}. \qquad (17)$$

This $m_t$-value agrees almost exactly with the central data value (16) (maybe too good).

The preliminary value (16) will be verified at LHC as a coming top-quark factory [10].

## 4. Geometric relation between the top-quark, neutrino and electron masses

The absolute QD-neutrino mass scale and the top-quark mass are special in particle physics since, among others, they define the beginning and the end of known elementary particle whole mass spectrum. Besides, neutrinos are special leptons[8], the top-mass is located close to the electroweak scale and is now experimentally the best known mass among quarks.

It is suggested by data that in relations (14) and (17) one needs to take into account the three-fold color mass-degeneracy of the top-quark. By analogy, neutrino-electron mass ratio in (11) takes into account three-fold neutrino quasi-degeneracy if we choose the appropriate coefficient a = $\pi/3$ (compare a = 0.96 – 1.53 in case of 3$\sigma$-ranges (8)),

$$3m_\nu/m_e = \pi \alpha_o^3, \qquad (18)$$

the coefficient '3' at the left side of (18) indicates sum of three QD-neutrino masses.

The connection $3m_\nu$-$m_e$ (18) is in oppositeness-relation to $3m_t$-$m_e$ one in (14). The physical meaning of the analogy between relations (14) and (18) for t-quark-electron and neutrino-electron mass ratios is that large deviations from mass-degeneracy of the two pairs (e, 3$\nu$) and (e, 3t) are quantitatively described by equal deviation from mass-degeneracy DMD-quantities: $[(m_e/3 m_\nu)-1] \cong [(3 m_t/ m_e)-1] \cong 1/(\pi \alpha_o^3)$.

The neutrino mass no longer appears separated from other particle masses; its minute mass is presented by small mass-ratio $\varepsilon$-deviation factor (from primary electron mass parameter $m_e$) as it is for all other particle masses.

Substantial implication of the semi-empirically suggested oppositeness relation between neutrino/electron and top-quark/electron mass ratios (18) and (14) is a simple geometric connection between the sums of three QD-

---

[8] Quasi-degenerate neutrinos are probably Majorana particles, but there is no decisive test yet.

neutrino masses ($3\,m_\nu$), three t-quark masses ($3\,m_t$) and single electron mass $m_e$:

$$(3\,m_\nu)(3\,m_t) = m_e^2. \qquad (19)$$

Unlike (11), (14) and (18), this interesting relation does not contain unknown parameters or coefficients. It closes the discrete sequence of elementary particle three-generation mass spectrum with singled out triplet of neutrino, top-quark and electron. Note that the equal coefficients 3 on the left side of (19) in the neutrino and t-quark terms have different physical meaning – the t-quark coefficient means three colors, but the neutrino one means three flavors.

Relation (19) is a *semi-empirical* geometric seesaw-like connection between neutrino and top-quark masses with electron mass $m_e$ at the geometric middle of the two extreme elementary particle masses.

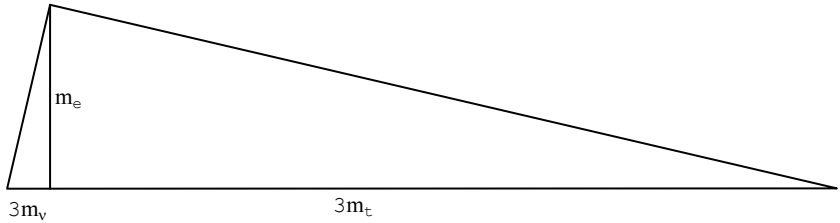

The smallest angles in the two similar right angled triangles with catheti measured by one pair of mass values ($m_e$, $3\,m_t$) and the other pair ($3\,m_\nu$, $m_e$) are equal and extremely small; with the top-quark central mass value from (16) we get:

$$\varphi = (3\,m_\nu)/m_e \cong m_e/(3\,m_t^{cntr}) \cong \pi\varepsilon^6(1+4\varepsilon^2) \cong 9.87 \times 10^{-7},$$

$$m_\nu \cong (\varphi/3)\, m_e, \quad m_t \cong m_e/3\varphi, \quad m_t/m_\nu \cong 1/\varphi^2 \cong 1.027 \times 10^{12}. \tag{20}$$

Top-quark mass $m_t$ in (19) is the large counterpart on the electroweak scale of the small neutrino mass. Such representation is possible only for QD-neutrinos and equal numbers of particle flavors and quark colors; the specifically neutrino condition of quasi-mass-degeneracy is related to the specifically quark condition of color mass-degeneracy.

Small neutrino mass versus large top-mass and reversely analogous phenomenon of large neutrino mixing versus small quark one are two distinct features of the way QD-neutrinos fit the pattern of elementary particle masses.

Note, approximate equality of the two very large mass ratios $(m_e/3m_\nu)$ and $(3m_t/m_e)$ in (19) and QD-neutrino type are experimentally falsifiable physical suggestions.

## 5. ε-hierarchical pattern of heavy quark masses

The considered semi-empirical approach to electron-particle mass ratios as small deviations (small ε–parameter) from their background values at $\varepsilon = 0$ (9) lead to the following estimations of three heavy quark masses:

$$m_e/3m_s \cong \pi\varepsilon^3, \quad m_s \cong 98 \text{ MeV}. \tag{21}$$

$$m_e/3m_c \cong \pi\varepsilon^4, \quad m_c \cong 1.19 \text{ GeV}, \tag{22}$$

$$m_e/3m_b \cong \pi^2\varepsilon^5, \quad m_b \cong 4.6 \text{ GeV}, \tag{23}$$

in agreement with PDG data [4].

Note some inferences,

$$2\pi(3m_s/m_\tau) \cong 1, \tag{24}$$

$$2\pi(3m_c/m_\mu)(m_e/m_\mu) \cong 1, \tag{25}$$

$$(\pi m_b)^2 \cong m_t m_c, \quad (26)$$

which have simple geometric interpretations.

All these relations, like those for t-quark (14), CL (2) and QD-neutrinos (18), are relations between sums of degenerate quark and lepton masses; obviously, the factor '3' appears explicitly only in quark-lepton or neutrino-CL relations, not in quark, CL or neutrino mass relations themselves.

To summarize, the semi-empirical mass ratios of electron mass $m_e$ to sums of color degenerate heavy quark masses $3m_k$ are approximately described by the $\varepsilon$-hierarchical pattern

$$m_e/3m_k \cong \pi \varepsilon^k, \quad m_e/3m_b \cong \pi^2 \varepsilon^5, \quad (27)$$

for the strange, charm and t-quark k = 3, 4, 6 respectively with an exception for the bottom quark where the important $\varepsilon$-power value is regular k = 5, but the coefficient is $\sim \pi^2$ instead of regular $\pi$. That b-quark exception cannot be embraced by the discussed semi-empirical phenomenology.

The main result from the semi-empirical quest for particle flavor pattern can be stated as: all considered elementary particle bare pole masses are expressed through the electron mass and discrete integer powers n = 1 ÷ 6 of the universal flavor parameter $\varepsilon$.

## 6. Absolute QD-neutrino mass scale

Compare the obtained from (18) value of neutrino mass scale

$$m_\nu = [\pi \alpha_o^3/3)m_e \cong 0.164 \text{ eV} \quad (18')$$

with direct estimation of neutrino mass from the PDG experimental top-quark mass data $(m_t)_{exp}$ (12), electron mass $m_e$ and geometric relation (19):

$$m_\nu = (m_e)^2/9(m_t)_{exp} = (0.167 \pm 0.003) \text{ eV}. \qquad (28)$$

The agreement between these two estimations of neutrino mass scale is to within 1 S.D.

Estimations (18) and (28) agree with the previous ones (7) and (8) based on totally independent neutrino oscillation data.

Both neutrino mass estimations (28) and (18') are reasonably fitting the discussed in the literature, see e.g. [7, 9] and [4, 6], astrophysical and terrestrial experimental constraints for the neutrino masses.

From relation (17) for $m_e/3m_t$ and oppositeness-connection between neutrino and t-quark masses (19) the average neutrino mass is given by

$$m_\nu = (\pi \alpha_o^3/3)(1+4\alpha_o)m_e \cong 0.168 \text{ eV}, \qquad (29)$$

in agreement with the central $m_\nu$-value from new top-quark experimental data (16) and geometric relation (19),

$$m_\nu = m_e^2/9(m_t)_{exp} = (0.168 \pm 0.001) \text{ eV}. \qquad (30)$$

The sum of three neutrino masses is

$$\Sigma m_\nu = (0.504 \pm 0.003) \text{ eV}, \qquad (31)$$

it fits the astrophysical constraints [9]:

$$(\Sigma m_\nu)_{exp} < 0.61 \text{ eV } (95\% \text{ CL}). \qquad (32)$$

Relation (18) for neutrino-electron mass-ratio means that the considered in [3] quadratic hierarchy for CL triplet τ-μ-e mass ratios,

$$(m_e/m_\mu) \cong \sqrt{2}\,(m_\mu/m_\tau)^2, \qquad (33)$$

transforms into a more rapid cubic one in the direction of lower lepton masses, i.e. for the μ-e-ν triplet, with QD-neutrino mass $m_\nu$ were the fourths lepton mass level. That cubic μ-e-ν mass ratio hierarchy is given by

$$(3m_\nu/m_e) \cong \sqrt{2}\,[\,2\pi\,(m_e/m_\mu)^3\,]. \qquad (34)$$

Relations (33) and (34) mean a pattern of the entire lepton mass spectrum that quantitatively defines the relatively tiny neutrino mass by steps τ–μ–e–ν down.

## 7. Relations between 3-flavor and electroweak basic physical quantities

A) Factual connection between the flavor ε-parameter and fine structure constant

The small universal ε-parameter (1) plays a crucial role in the discussed semi-empirical particle flavor physics phenomenology; it determines a *pattern* of particle-electron mass ratios and large versus small neutrino and quark mixing angles in noticeable agreement with data – very unlikely to be a *system* of coincidences. Probably, this parameter has basic meaning.

The old flavor physical problem of electron-muon mass ratio and its closeness to the fine structure constant does not have theoretical or phenomenological explanation today as about half a century ago despite the achieved highly accurate experimental data; attempts to address it are welcome [14].

I consider hints from experimental data that point to relevant phenomenological ideas. The constant $\alpha_o = \varepsilon^2$ is close (~92%) to the fine structure constant $\alpha$ at zero momentum transfer. And there exists an almost exact factual connection between the constants $\alpha_o$ and $\alpha$ [15],

$$(\exp\alpha/\alpha)^{\exp 2\alpha} + (\alpha/\pi) = 1/\varepsilon^2. \qquad (42)$$

This relation is accurate to within $(\alpha - \alpha_{Data})/\alpha_{Data} \cong 10^{-8}$ with $\alpha_{Data}$ from [4], [16]. Equation (42) is extendable,

$$(\exp\alpha/\alpha)^{\exp 2\alpha} + [(\alpha/\pi) + O(\alpha^2)] = 1/\varepsilon^2. \qquad (43)$$

The first term on the left side includes nonperturbative corrections to $(\varepsilon^2-\alpha)$-connection, while the second term should include all perturbative ones[9]. In fact $O(\alpha^2) \cong -\alpha^2/4\pi$ to within an accuracy $10^{-10}$ [15]. Since the value $\varepsilon$ is real by definition, from Eq.(43) follows that $\alpha$ is a positive-definite solution as it must be for the fine structure constant.

The universal parameter $\varepsilon$ determines all considered above bare particle flavor quantities as if it is related to a dynamical constant; the factual highly accurate equation (43) describes that relation.

As a result of (2) and (43), approximate relations for CL bare mass ratios follow

$$(m_e/m_\mu) \cong \varepsilon^2/\sqrt{2}, \quad (m_\mu/m_\tau) \cong \varepsilon/\sqrt{2},$$

$$(\varepsilon - \sqrt{\alpha})/\sqrt{\alpha} \cong -0.04, \quad \varepsilon \cong 0.96\sqrt{\alpha} = 0.96(e/\sqrt{4\pi}), \qquad (44)$$

where 'e' is the absolute electric charge of the electron.

The approximate relations (44) between electron-muon and muon-tau mass ratios and the fine structure constant $\alpha(q^2=0)$ – *most importantly since they are a regular part of the observed elementary particle general flavor ε-pattern* – present a phenomenological explanation[10],[11] of the mentioned above old $(\mu/e)$-alpha problem.

---

[9] The small empirical perturbative corrections are represented here by powers of the small quantity $\alpha$, while the empirical nonperturbative corrections - by powers of the unique order-one quantity $\exp\alpha$. Both are finite corrections, comp. footnote-5.

[10] Commonly, the main point of a useful primary phenomenological explanation of a basic problem (preceding deeper theoretical explanation) is to show that the proposed particular description of that problem is related to other known unsolved or already solved ones by a *definite pattern.*

The above discussion suggests that particle 3-flavor physics and the highly successful one-generation electroweak physics are essentially connected through the universal small parameter ε. If ε → 0, the considered ratios of second and third generation particle masses to the electron mass parameter would increase infinitely, and the extra particle generations would be unobservable. But at the same time the electroweak interactions and atomic bound states of the first generation particles would vanish: α → $α_o$ → 0. Therefore, the above discussion points to a substantially united 3-flavor-electroweak phenomenology: in its framework there would be no atoms if there are not two observable extra elementary particle generations (the 'muon' and 'tau' ones) beyond the first ('electron') particle generation.

B) On physical meaning of the neutrino oscillation solar-atmospheric hierarchy parameter

DMD-quantities and hierarchies of CL and neutrinos are basic dimensionless observable quantities in lepton flavor physics. There are two DMD-quantities and one DMD-hierarchy for *three* charged leptons and same for the *three* neutrinos:

DMD(CL)1 = $[(m_\tau^2/m_\mu^2)-1]$, DMD(CL)2 = $[(m_\mu^2/m_e^2)-1]$,

$\quad$ h-DMD(CL) = DMD(CL)1 / DMD(CL)2;  $\qquad$ (45)

DMD(ν)1 = $[(m_2^2/m_1^2)-1]$, DMD(ν)2 = $[(m_3^2/m_2^2)-1]$,

$\quad$ h-DMD(ν) = DMD(ν)1 / DMD(ν)2.  $\qquad$ (46)

By definition, the mass-squared DMD-hierarchy (h-DMD) quantity of CL from (2) is given by

$$\text{h-DMD(CL)} \cong \varepsilon^2 \cong \alpha, \tag{47}$$

it is close to the fine structure constant $\alpha$ at pole value of the photon propagator $q^2 = 0$.

---

[11] There are also other approximate nontrivial relations between particle mass ratios, mixing angles and the fine structure constant (e.g. (5),(6)) that attract the 'why'-attention, but more often get suppressed.

The DMD-hierarchy quantity of QD-neutrinos from (3) and (4) is given by

$$\text{h-DMD}(\nu) \cong r \cong 5\alpha_o \cong 0.03369, \tag{48}$$

it is equal to the solar-atmospheric hierarchy parameter $r$ from the neutrino oscillation experimental data and its numerical value is close to $5\alpha_o$. On the other hand, the quantity $5\alpha_o$ is close to the semi-weak analog $\alpha_W$ of the fine structure constant $\alpha$ at pole value of the Z-boson propagator from PDG data [4]:

$$\alpha(M_Z) = 1/(128.91 \pm 0.02), \quad (\sin^2\theta_W)|_{M_Z} = 0.23108 \pm 0.00005,$$
$$\alpha_W(M_Z) = (\alpha/\sin^2\theta_W)|_{M_Z} \cong 0.03357. \tag{49}$$

The numbers on the right in (48) and (49), i.e. in the relation $\alpha_W(M_Z) \cong 5\alpha_o$, disagree by less than 1%.

The best-fit values and $3\sigma$ allowed ranges for solar-atmospheric hierarchy parameter from the three-flavor neutrino oscillation global data are given by [17]

$$r_{b.f.} = 0.032, \quad 0.027 \leq r_{3\sigma} \leq 0.040. \tag{50}$$

The suggested quantitative connection between oscillation hierarchy parameter $r$ and flavor parameter $\varepsilon$

$$r \cong -\varepsilon^2 \log \varepsilon^2 \cong 0.03369 \tag{51}$$

is, from (50), well within the $3\sigma$ ranges and agrees with the best fit value $r_{b.f.}$ within accuracy 5%.

If further confirmed by experimental data with higher accuracy and confidence for both the $r$-parameter at neutrino oscillation experiments and for the electroweak interaction constant $\alpha_W(q^2 = M_Z^2)$ at Z-pole experiments, the relation (48) between neutrino DMD-hierarchy, oscillation hierarchy parameter and the semi-weak dynamical electroweak constant $\alpha_W$,

$$\text{h-DMD}(\nu) = r \cong \alpha_W(M_Z), \qquad (52)$$

will be an informative evidence in favor of QD-neutrino type as from new fundamental physics.

Relation (52) may be the true physical meaning of the neutrino oscillation solar-atmospheric hierarchy parameter $r$. This relation hints on a quantitative answer of why is the solar neutrino mass-squared difference much smaller than the atmospheric one. In that regard it is interesting to address the case of not-QD-neutrinos. In general with the choice $m_1 < m_2 < m_3$ it follows

$$\text{h-DMD}(\nu) = (m_1^2/m_2^2)r. \qquad (53)$$

If the physical meaning of the solar-atmospheric hierarchy parameter $r$ is indeed the unique DMD-hierarchy h-DMD($\nu$), the condition

$$(m_1^2/m_2^2) \cong 1 \qquad (54)$$

must be fulfilled. In case of inverse neutrino mass ordering ('hierarchy') the condition (54) means QD-neutrinos. But there is a special case of not-QD-neutrinos - with 'normal' neutrino mass ordering and relations[12]

$$\Delta m^2_{sol} << m_1^2, m_2^2 < \approx \Delta m^2_{atm}, \qquad (55)$$

where the condition h-DMD($\nu$) $\cong r$ may be also approximately fulfilled. Only in this particular not-QD-neutrino case the solar-atmospheric hierarchy parameter $r$ may have the more general[13] physical meaning of hierarchy of deviations from mass-degeneracy between the two pairs of neutrino masses $(m_2, m_1)$ and $(m_2, m_3)$,

$$\text{h-DMD}(\nu_{N\text{-}QD}) = [(m_2^2/m_1^2)-1]/[(m_3^2/m_2^2)-1]\big|_{(54)} \cong r, \quad (56)$$

as it is in the QD-case. But in the not-QD case relation (56) seems an accidental one (normal ordering and restricted neutrino mass interval as tuning conditions) in contrast to QD-neutrino case where it is fulfilled by the QD-definition. In any case, the importance of further verification of that new physics informative relation (52) between the neutrino DMD-hierarchy, solar-atmospheric parameter $r$ and electroweak constant $\alpha_W(M_Z)$ cannot be overestimated.

---

[12] For example $m_2^2 \cong m_1^2 \cong (0.1 \div 5) \times 10^{-3}$ eV$^2$, $\Delta m^2_{atm} \cong 2.4 \times 10^{-3}$ eV$^2$ [17].

[13] Unlike the primary definition of the solar-atmospheric hierarchy parameter $r$, DMD-hierarchy is a general physical notion applicable also to 3-flavor mass systems such as CL and quarks.

To summarize, neutrino mass-degeneracy deviations are two dimensionless quantities with definite meaning in neutrino physics in contrast to the two mass-squared differences, which magnitudes have not independent (small or large) physical meaning. The true physical meaning of the QD-neutrino oscillation hierarchy-parameter *r as DMD-hierarchy quantity* (48) (that is independent of the 'solar-atmospheric' particulars) – $r \cong$ h-DMD($\nu$) - enables its possible relation to the electroweak interaction constant $\alpha_W(M_Z)$ in analogy with the CL relation (47) h-DMD(CL) $\cong \alpha$.

## 8. Conclusions

1) In contrast to the one-generation Standard Model, an established theory of particle flavor does not exist today. In this paper, tree approximation semi-empirical phenomenology of bare dimensionless flavor quantities is discussed. The considered bare values of particle pole mass ratios and mixing angles approximately agree with the available data. If they do, it means that the not calculable in known field theory physical radiative contributions to pole mass and dimensionless flavor quantities should be relatively small (comp. footnote-5).

2) The present research is pointing to elementary particle mass spectrum being a system of discrete mass levels dimensionally determined by the electron mass parameter $m_e$ and numerated by integer powers n of the ε-parameter $\varepsilon^n$, n = −1 ÷ −6 for quarks, n = −2, −3 for the charged leptons and n = +6 for QD-neutrinos.

3) In the established neutrino oscillation phenomenology the small magnitude of the solar-atmospheric hierarchy parameter *r* is an experimental fact without an explanation. As mentioned in Sec.1, in the QD-neutrino phenomenology the relations (3) definitely explain the smallness of that parameter $r \ll 1$, but the quantitative value $r \cong -\varepsilon^2 \log \varepsilon^2$ follows only from heuristic suggestion that the small parameter *r* be expressed through the universal parameter ε and data hints. The discussed in Sec.7 idea that unique (in case of three neutrino mass levels) DMD-hierarchy is the true physical meaning of the neutrino oscillation parameter *r* together with hints from the relevant experimental data suggests the connection $r \cong \alpha_W(M_Z) \cong 0.0336$, which may explain the real magnitude of that parameter. It indicates a deeper physical meaning of the neutrino oscillation solar-atmospheric hierarchy parameter *r* without reference to solar-atmospheric circumstantial particulars, and so proposes related only to neutrino physics name for that parameter – 'neutrino DMD-hierarchy' parameter.

4) The small flavor parameter ε is *universal* only in particle flavor phenomenology with quasi-degenerate neutrinos. It is quantitatively related to the fine structure constant α. The two DMD-hierarchy physical quantities of CL and QD-neutrinos are analogously close to the two electroweak coupling constants $\alpha$ and $\alpha_W$ at photon and Z-boson momentum transfers respectively. These data-hints point to an *essential* connection (through the universal $\varepsilon$–parameter) between the problematic flavor sector of particle physics and the electroweak sector of the known highly successful one-generation Standard Model theory.

 5) Realistic elementary particle flavor pattern of mass ratios and mixing angles is considered small deviated (measured by the small parameter $\varepsilon$) one from the stated 'background' flavor pattern. The latter is specified by electron mass $m_e$ as the sole finite mass parameter, one angle-parameter $\pi/4$, degenerate maximally mixed neutrinos and infinitely small neutrino/electron and electron/Dirac particles (muon, tau and quarks) mass ratios and minimally mixed quarks as represented by the definitions (9), (10). Since the actual particle masses are expressed through different powers of the dimensionless small $\varepsilon$-parameter, the very small (neutrino) and very large (top-quark) masses appear naturally – they differ from other particle masses of leptons and quarks only by values of the $\varepsilon$-powers.

  6) The found cubic in $\alpha_o$ and almost equal values for electron/top-quark and QD-neutrino/electron mass ratios $m_e/3m_t \cong 3m_\nu/m_e \cong \pi\alpha_o^3$ mean dual relation between these two extreme mass ratios and equally large deviations ($\cong 1/\pi\alpha_o^3$) from degeneracy between the electron mass and the two mass sums of suggested QD-neutrinos and color-degenerate top-quark.

  7) Independent of any parameters geometric relation for sums of three-fold degenerate neutrino and top-quark masses $(3\,m_\nu) = m_e^2/(3\,m_t)$ is suggested. It is a seesaw-like connection between the very different QD-neutrino and t-quark masses with large top-quark mass $m_n$ as the opposite counterpart of the small neutrino mass $m_\nu$. Dual relation between neutrino-electron and top-quark-electron mass ratios is a quantitative prediction of QD-neutrino type independent of suggestions of Dirac or Majorana neutrino nature. Experimental confirmation of that $(3 m_\nu - m_e - 3 m_t)$-relation means discovering QD-neutrino type with $m_\nu \cong 10^{-12} m_t \cong 0.17$ eV.

8) The obtained sub-eV value $\Sigma m_\nu = (0.50 \pm 0.003)$ eV for the sum of QD-neutrino masses follows from experimental top-quark mass data and is in agreement with the previous neutrino mass estimations (7), (8) from independent experimental data on oscillation mass squared differences. That sub-eV QD-neutrino mass sum is compatible with known cosmological and terrestrial experimental constraints on neutrino masses, and be tested by coming data.